\documentclass[twocolumn,showpacs,preprintnumbers]{revtex4} 
\input{epsf}

\def\AJ{{\it Astroph. J.} }
\def\AJL{{\it Ap. J. Lett.} }

\def\ASJ{{\it Astron. J.} }

\def\CQG{{\it Class. Quantum Gravity} }

\def\FP{{\it Fortschr. Physik} }
\def\GRG{{\it Gen. Relativity and Gravitation} }

\def\JHEP{{\it JHEP} }

\def\MPL{{\it Mod. Phys. Lett.} }

\def\NAT{{\it Nature} }
 
\def\NC{{\it Il Nuovo Cimento} }
\def\NP{{\it Nucl. Phys.} }
\def\PL{{\it Phys. Lett.} }
\def\PR{{\it Phys. Rev.} }
\def\PRL{{\it Phys. Rev. Lett.} }

\def\frac#1#2{{\textstyle{{#1}\over {#2}}}}

\def\lsim{\mathrel{\rlap{\lower4pt\hbox{\hskip1pt$\sim$}}
    \raise1pt\hbox{$<$}}}
\def\gsim{\mathrel{\rlap{\lower4pt\hbox{\hskip1pt$\sim$}}
    \raise1pt\hbox{$>$}}}
\def\sqr#1#2{{\vcenter{\vbox{\hrule height.#2pt
         \hbox{\vrule width.#2pt height#1pt \kern#1pt
         \vrule width.#2pt}
         \hrule height.#2pt}}}}

\def\beq{\begin{equation}}
\def\eeq{\end{equation}}
\def\beqa{\begin{eqnarray}} 
\def\eeqa{\end{eqnarray}}

\begin{document}


\title{Generalized Chaplygin gas and CMBR constraints}

\author{M. C. Bento$^{1,2}$, O. Bertolami$^1$ and A.A. Sen$^3$}

\affiliation{$^1$  Departamento de F\'\i sica, Instituto Superior T\'ecnico \\
Av. Rovisco Pais 1, 1049-001 Lisboa, Portugal}

\affiliation{$^2$  Centro de F\'{\i}sica das 
Interac\c c\~oes Fundamentais, Instituto Superior T\'ecnico}

\affiliation{$^3$  Centro Multidisciplinar de Astrof\'{\i}sica, 
Instituto Superior T\'ecnico}

\affiliation{E-mail addresses: bento@sirius.ist.utl.pt; orfeu@cosmos.ist.utl.pt; 
anjan@x9.ist.utl.pt}

\vskip 0.2cm

\date{\today}


\begin{abstract}
     
We study the dependence of the location  of the Cosmic Microwave
Background Radiation (CMBR) peaks  on the parameters of the generalized
Chaplygin gas  model, whose equation of state is given by
$p = - A/\rho^{\alpha}$, where $A$ 
is a positive constant and $0 < \alpha \le 1$. We find, in particular, 
that observational data arising from Archeops for the location of the
first peak, BOOMERANG for the location of the
third peak, supernova and high-redshift observations 
allow constraining significantly the  parameter space of the model.
Our analysis indicates that the emerging model is clearly distinguishable 
from the $\alpha = 1$ Chaplygin case and the $\Lambda$CDM model. 
 
\vskip 0.2cm
 
\end{abstract}

\pacs{ 98.80.Cq, 98.65.Es \hspace{2cm}Preprint DF/IST-10.2002}

\maketitle
\section{Introduction}

It has been recently suggested that the change of behaviour of the so-called  
dark energy density might be controlled by the change in the equation 
of state of the background fluid \cite{Kamenshchik} 
instead of the form of the potential, 
thereby avoiding well known fine-tuning problems of quintessence models.
This is achieved via the introduction, within the framework of 
Friedmann-Robertson-Walker cosmology, 
of an exotic background fluid, the generalized Chaplygin gas, 
described by the equation of state

\beq
p_{ch} = - {A \over \rho_{ch}^\alpha}~~,
\label{eq:eqstate}
\eeq
where $\alpha$ is a constant in the range 
$0 < \alpha \le 1$ (the Chaplygin gas corresponds to the case $\alpha=1$) and
$A$ a positive constant. Inserting this equation of state 
into the relativistic energy conservation equation, leads to a density evolving
as \cite{Bento1}

\beq
\rho_{ch} =  \left(A + {B \over a^{3 (1 + \alpha)}}\right)^{1 \over 1 + \alpha}~~,
\label{eq:rhoc}
\eeq 
where $a$ is the scale factor of the Universe and $B$ an integration 
constant. Remarkably, the model  interpolates between 
a universe dominated by dust and a De Sitter one via a phase described 
by a ``soft'' matter equation of state, $p = \alpha \rho$ ($\alpha \not= 1$).
Notice that even though Eq. (\ref{eq:eqstate}) admits a wider range of
positive $\alpha$ values, the chosen range of values ensures that 
the sound velocity ($c_s^2 = \alpha A/ \rho_{ch}^{1+\alpha}$) does not exceed,
in the ``soft'' equation of state phase, 
the velocity of light. Actually, as discussed in Ref. \cite{Bento1}, 
it is only for values in the range $0 < \alpha \le 1$ 
that the analysis of the evolution of energy density fluctuations makes sense.
  
It was also shown in Ref. \cite{Bento1} that the model can be described by a
complex scalar field whose action can be written as a generalized Born-Infeld
action corresponding to  a  ``perturbed'' $d$-brane in a $(d+1, 1)$ spacetime. 
It is clear that this model has a bearing on the observed
accelerated expansion of the Universe \cite{Perlmutter} as it automatically 
leads to an asymptotic phase where the equation of state is dominated by a 
cosmological constant, $8 \pi G A^{1/1+\alpha}$. 
It was also shown that the model admits, under conditions, 
an inhomogeneous generalization which can be regarded as a unification 
of dark matter and dark energy \cite{Bilic,Bento1} and that it  can
be accomodated within the standard structure formation
scenarios \cite{Bento1,Bilic,Fabris}.  Therefore, the  generalized Chaplygin
gas model seems to be a viable alternative to models where the accelerated
expansion of the Universe is 
explained through an uncancelled cosmological constant (see \cite{Bento2} and 
references therein) or through quintessence models with one 
\cite{Bronstein,Ratra,Wetterich,Caldwell,Ferreira,Zlatev,Binetruy,Kim,Uzan,
Skordis} 
or two scalar fields \cite{Fujii,Masiero,Bento3}.  

These promising results have led, quite recently, to a wave of interest aiming 
to constrain the generalized Chaplygin model using  observational data, 
particularly those  arising
from SNe Ia \cite{Fabris1,Avelino,Dev,Gorini,Makler,Alcaniz}.

In this work, we shall consider the constraints arising from the positions of
the first three CMBR peaks on the parameter space of the generalized
Chaplygin gas, applying the same method that has been used recently 
to constrain quintessence models (see e.g. Refs.\cite{Doran1,Doran2,Domenico}). 
We find, in particular, that the 
positions of first and third peaks lead to fairly strong constraints
although a sizeable portion of the parameter space of the model is still compatible
with BOOMERANG and Archeops data. Further correlating the resulting region with 
the observations of supernova and high-redshift objects leads to quite tight 
contraint on the parameter space of the generalized Chaplygin model.  
It is important to stress that the generalized Chaplygin gas differs, as discussed
in Ref. \cite{Gorini} from quintessence and tracker
models in what concerns the so-called ``statefinder'' parameters $(r, s)$ 
\cite{Sahni}:

\beq 
r \equiv {\dddot{a} \over a H^3} ~~~,~~~ s \equiv {r -1 \over 3 (q - 1/2)}
\label{eq:statefinder}
\eeq 
where $H = \dot{a} /a$ is the Hubble parameter and $q = - \ddot{a}/aH^2$ is the 
deceleration parameter. Moreover, only for fairly small values of $\alpha$ 
the generalized Chaplygin gas becomes indistinguishable from the $\Lambda$CDM.
Hence, future SNe Ia surveys for high redshifts may allow a clear 
discrimination between the generalised Chaplygin gas proposal and
quintessence/tracking models. 

\begin{figure}[t]
\centering
\leavevmode \epsfysize=7cm \epsfbox{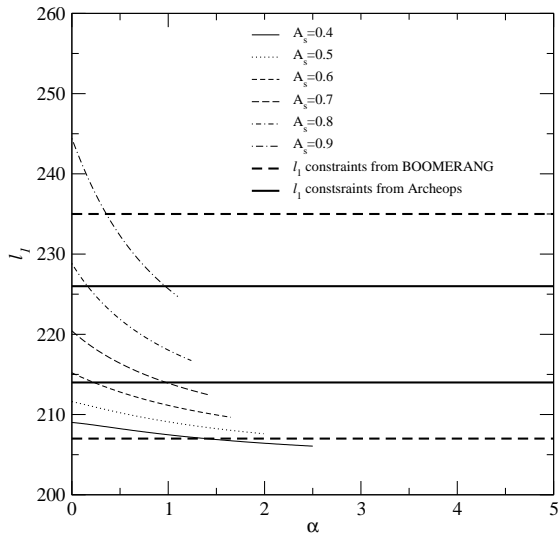}\\
\vskip 0.1cm
\caption{Dependence of the position of the CMBR first peak, $l_1$, 
as a function of $\alpha$ for different values of  $A_S$. Also shown are
the observational bounds on $l_1$ from BOOMERANG (dashed lines), 
see Eq.~(\ref{eq:l1BOOM}), and Archeops (full lines), see Eq.~(\ref{eq:l1Arch}).}
\label{l1}
\end{figure}

\section{Location of CMBR peaks for the generalized Chaplygin gas }

The CMBR peaks arise from acoustic oscillations of the primeval plasma just
before the 
Universe becomes transparent. The angular momentum scale of the oscillations
is set by the acoustic scale $l_A$ which for a flat Universe is given by

\beq
\label{eq:la}
l_A = \pi {\tau_0 - \tau_{\rm ls} \over \bar c_s \tau_{\rm ls}}~~,
\eeq
where $\tau_0$ and $\tau_{\rm ls}$ are the conformal time today and at
last scattering and $\bar{c}_s$ is the average sound speed before decoupling.

The prior assumptions in our subsequent calculations are as follows:  
scale factor at present $a_{0} = 1$, scale factor at last scattering 
$a_{ls} = 1100^{-1}$, $h = 0.65$, density parameter for radiation and baryons
at present $\Omega_{r0} = 9.89 \times 10^{-5}$, 
$\Omega_{b0} = 0.05$, average sound velocity $\bar{c}_{s} = 0.52$, 
and spectral index for the initial energy density perturbations, 
$n = 1$
 
We start by computing $l_A$ for the case of the generalized Chaplygin gas.
Rewriting Eq.~(\ref{eq:rhoc}) in the form
\beq
\label{eq:rho_c0}
\rho_{ch}=\rho_{ch0}\left( A_s + {(1-A_s)\over
a^{3(1+\alpha)}}\right)^{1/1+\alpha}~~,
\eeq
where $A_s\equiv A/ \rho_{ch0}^{1+\alpha}$ and
$\rho_{ch0}=(A+B)^{1/ 1+\alpha}$, 
the Friedmann equation becomes

\beq
\label{eq:H2}
H^2={8\pi G\over 3}\left[{\rho_{r0}\over a^4}+{\rho_{b0}\over{a^3}}+\rho_{ch0}
\left( A_s + 
{(1-A_s)\over a^{3(1+\alpha)}}\right)^{1/1+\alpha}\right]~~,
\eeq
where we have included the contribution of radiation and baryons as this is not
accounted for by the generalized Chaplygin gas equation of state. 

Several important features of  Eq. (\ref{eq:rho_c0}) are worth remarking. 
First of all, $A_s$ must lie in the interval $0 \leq A_{s} \leq 1$ as 
otherwise $p_{ch}$ will be undefined at some $a$. Secondly, for $A_{s} = 0$,
the Chaplygin gas behaves as  dust and, for $A_{s} = 1$, it behaves like as a 
cosmological constant. Notice that only  for $\alpha = 0$,  the Chaplygin gas
corresponds to a $\Lambda$CDM model. 
Hence, for the chosen range of $\alpha$, the generalised Chaplygin 
gas is clearly different from $\Lambda$CDM. Another relevant issue is that
the sound velocity of the fluid  is given, at present, by $\alpha A_s$ and
thus $\alpha A_s \le 1$. Using

\beq
{\rho_{r0}\over \rho_{ch0}}={\Omega_{r0}\over\Omega_{ch0}}=
{\Omega_{r0}\over 1-\Omega_{r0}-\Omega_{b0}},
\eeq

\noindent
and

\beq
{\rho_{b0}\over \rho_{ch0}}={\Omega_{b0}\over\Omega_{ch0}}=
{\Omega_{b0}\over 1-\Omega_{r0}-\Omega_{b0}},
\eeq

\noindent
we obtain

\beq
\label{eq:H2final}
H^2=\Omega_{ch0} H_0^2 a^{-4} X^2(a)~~,
\eeq
with

\beqa
\label{eq:defX}
X(a) &=& {\Omega_{r0}\over 1-\Omega_{r0}-\Omega_{b0}} + 
{\Omega_{b0}~a \over 1-\Omega_{r0}-\Omega_{b0}} \nonumber\\
&+&a^4 \left( A_s + {(1-A_s)\over
a^{3(1+\alpha)}}\right)^{1/1+\alpha}~~.
\eeqa
Using the fact that $H^2=a^{-4} \left(d a\over d \tau\right)^2$, we get

\beq
\label{eq:dtau}
d\tau={da\over \Omega_{ch0}^{1/2} H_0 X(a)}~~,
\eeq
so that

\beq
\label{eq:lA}
l_A={\pi\over \bar c_s}\left[{\int_0}^1 {da\over X(a)} \left( {\int_0}^{a_{ls}}
{da\over X(a)}\right)^{-1} - 1 \right]~~.
\eeq

\begin{figure}[t]
\centering
\leavevmode \epsfysize=7cm \epsfbox{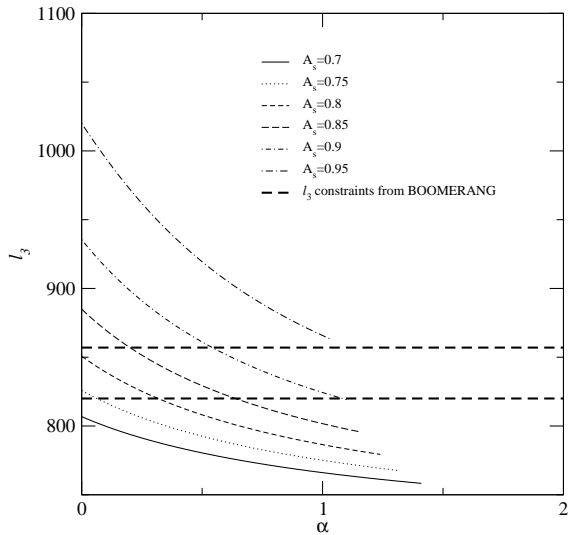}\\
\vskip 0.1cm
\caption{Dependence of the position of the CMBR third peak, $l_3$, as
a function of $\alpha$ for different values of $A_S$. Also shown are
the observational  bounds on  $l_3$ (dashed lines), see Eq.~(\ref{eq:l3obs}).}
\label{l3}
\end{figure}

In an idealised model of the primeval plasma, there is a simple relation
between the location of the $m$-th peak and the acoustic scale, namely
$l_m\approx m l_A$. However, the location of the peaks is slightly shifted
by driving effects and
this can be compensated by  parameterising the location of the $m$-th
peak, $l_m$, as in \cite{Hu,Doran1}

\beq 
\label{eq:lm}
l_m \equiv l_A \left(m - \varphi_m\right)~~. 
\eeq

It is not in general possible to derive analytically
a relationship between the cosmological parameters and the peak shifts,
but one can use fitting formulae that describe their dependence on these
parameters; in particular, we have for the spectral index of scalar
perturbations $n=1$ and for the amount of baryons $\Omega_{b0} h^2=0.02$
\cite{Hu,Doran1}

\beq
\label{eq:varphi1}
\varphi_1\approx 0.267 \left({r_{ls} \over 0.3}\right)^{0.1}~~,
\eeq
where $r_{ls}=\rho_r(z_{ls})/\rho_m(z_{ls})$ is the ratio of radiation
to matter at last scattering. Since, according to our dark matter-energy
unification hypothesis,  $\rho_{ch}$ will behave as dust or non-relativistic 
matter at last scattering

\beq
\label{eq:rhocls}
\rho_{ch}\approx {\rho_{ch0}\over a^3} (1-A_s)^{1/1+\alpha}~~, 
\eeq
we get

\beqa
\label{eq:rls}
r_{ls}&= &{\Omega_{r0}\over \Omega_{ch0}} {a_{ls}^{-1}\over
 (1-A_s)^{1/1+\alpha}} \nonumber\\
&\simeq & {\Omega_{r0} a_{ls}^{-1}\over(1- \Omega_{r0}-\Omega_{b0})
 (1-A_s)^{1/1+\alpha}}~~.
\eeqa
Using Eqs. (\ref{eq:lA}) and  (\ref{eq:lm} )-(\ref{eq:rls}), we have
plotted in Figure~1, $l_1$ as a function of $\alpha$ for different values
 of $A_s$, 
where we have also drawn lines corresponding to the observational 
bounds on $l_1$ as derived from BOOMERANG \cite{Boomerang} (dashed lines)

\beq
\label{eq:l1BOOM}
l_1  =  221\pm 14~~.
\eeq
and Archeops data \cite{Archeops} (full lines)

\beq
\label{eq:l1Arch}
l_1  =  220\pm 6~~.
\eeq
Notice that, since  $\alpha A_s\leq 1$, for a specific value of $A_s$ curves
end where this relation gets saturated, $\alpha A_s = 1$.

It is very difficult to extract any constraints from the position of the 
second peak since it depends on too many parameters, hence we shall disregard
it hereafter.

As for the shift of the third peak, it turns out to be a relatively
insensitive quantity \cite{Doran2}

\beq
\label{eq:phi3}
 \varphi_3\approx 0.341~~.
\eeq
Figure~2 shows $l_3$ as a function of $\alpha$ for different values of $A_s$,
where the dashed lines are the current lower and upper bounds
on $l_3$ as derived from BOOMERANG data \cite{Boomerang} 

\beq
\label{eq:l3obs}
l_3  =  845_{-25}^{+12}~~.
\eeq

We see that $l_3$ puts rather  tight constraints on the
parameters of the model, $\alpha$ and $A_s$. 

Figure~3 shows the constraints on the parameter space of the generalized
Chaplygin gas model, the ($A_s, \alpha$) plane, that are obtained from the
observational bounds on the location of the first (full contour) and third (dashed contour)
CMBR peaks. Hence, from the CMBR point of view 
the allowed region of the model parameters lies in the 
intersection between these two contours.

\begin{figure}[t]
\centering
\leavevmode \epsfysize=7cm \epsfbox{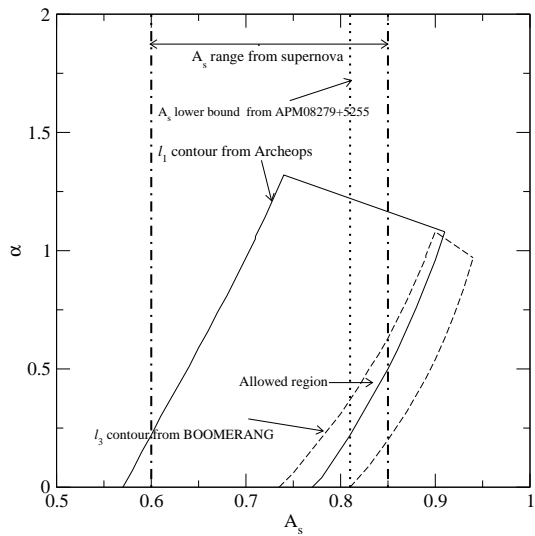}\\
\vskip 0.1cm
\caption{Contours in the ($\alpha$, $A_S$) plane arising from Archeops constraints 
on $l_1$ (full contour) and  BOOMERANG constraints on $l_3$ (dashed contour), supernova 
and APM $08279+5255$ object. 
The allowed region of the model parameters lies in the intersection between 
these regions.}
\label{param2}
\end{figure}

\section{Discussion and Conclusions}

In this paper we have shown that the location of the CMBR peaks,
as determined  via Archeops and BOOMERANG data, allows constraining a 
sizeable portion
of the parameter space of the generalized Chaplygin gas model. Our results indicate
that the constraints arising from the position of the first peak, as recently
announced by the Archeops collaboration, imply, for $\alpha \le 1$, that
$0.57 \lsim A_s \lsim 0.91$.
 
On the other hand, the location of the third acoustic peak arising from the 
BOOMERANG collaboration provides 
strong constraints on the parameter space of the model,
as indicated in Figure 3 (dashed contour region).
Notice that compatibility with data requires that only the fairly small 
intersecting region is allowed, that is $0.74 \lsim A_s \lsim 0.90$; 
consistency with SNe Ia data suggests on its hand that $0.6 \lsim A_s \lsim 0.85$ 
\cite{Makler}, and this together with the bound arisig from the 
APM $08279+5255$ source, $A_s \ge 0.81$, \cite{Alcaniz} lead us to 
obtain a fairly tight constraint $0.81 \lsim A_s \lsim 0.85$ and 
$0.2 \lsim \alpha \lsim 0.6$. 
Furthermore, we stress that the allowed region in Figure 3
is clearly distinct from the Chalpygin gas ($\alpha = 1$) and the 
$\Lambda$CDM model. 

Clearly, with future high precision measurements of the MAP and PLANCK
satellites, we expect that the position of the first three peaks will
be determined to high accuracy, thus allowing further constraints on 
the parameter space of the generalized Chaplygin gas model. Correlating the
resulting constraints with SNe Ia, red-shift objects and, for instance,  
gravitational lensing
data  may uniquely determine these parameters.

\vskip 0.3cm

\centerline{\bf {Acknowledgments}}

\vskip 0.2cm

\noindent
A.A.S. is grateful to T. Barreiro for useful discussions. M.C.B. and  O.B.
acknowledge the partial support of Funda\c c\~ao para a 
Ci\^encia e a Tecnologia (Portugal)
under the grant POCTI/1999/FIS/36285. The work of A.A.S. is fully 
financed by the same grant. 


\end{document}